\documentclass[runningheads]{llncs}
\newcommand{\longversion}[1]{#1}
\newcommand{\shortversion}[1]{}
\usepackage{todonotes}
\usepackage{comment}

\usepackage[utf8]{inputenc}          
\usepackage[T1]{fontenc}             
\usepackage[USenglish]{babel}          

\usepackage[intlimits]{amsmath}      
\usepackage{amsfonts}                
\usepackage{amssymb}                 
\usepackage{amsmath}                 
\allowdisplaybreaks
\shortversion{\usepackage{apxproof}
} 
\longversion{\usepackage[appendix=inline]{apxproof}}
\shortversion{\usepackage{apxproof}}
\usepackage{mathtools}
\usepackage{setspace}
\usepackage{graphicx}
\usepackage{hyperref}
\usepackage{xcolor}
\usepackage{xspace}
\usepackage{amsthm}
\newtheorem{observation}{Observation}
\usepackage{caption}
\usepackage{subcaption}
\usepackage{tikz}
\usetikzlibrary{shapes.geometric}

\usepackage{algorithm}
\usepackage[noend]{algpseudocode}

\newcommand{\snd}{\textsf{snd}}

\newcommand{\vc}{\textsf{vc}}
\newcommand{\tw}{\textsf{tw}}
\newcommand{\dtw}{\textsf{dtw}}
\newcommand{\fvs}{\textsf{fvs}}
\newcommand{\pw}{\textsf{pw}}
\newcommand{\td}{\textsf{td}}
\newcommand{\NP}{\textsf{NP}}
\newcommand{\FPT}{\textsf{FPT}\xspace}

\newcommand{\W}[1]{\ensuremath{\textsf{W}[#1]}}

\newcommand{\iffl}{if\longversion{ and only i}f }

\newcommand{\yes}{\textsf{yes}}

\newenvironment{pf}{\begin{proof}}{\hfill\qed\end{proof}}
\newenvironment{pfclaim}{\begin{proof}}{\hfill$\Diamond$\end{proof}}

\makeatletter
\newcommand{\crossout}[1]{%
  \begingroup
  \sbox\z@{#1}%
  \dimen\z@=\wd\z@
  \dimen\tw@=\ht\z@
  \dimen\z@=.99626\dimen\z@   
  \dimen\tw@=.99626\dimen\tw@ 
  \edef\co@wd{\strip@pt\dimen\z@}
  \edef\co@ht{\strip@pt\dimen\tw@}
  \leavevmode
  \rlap{\pdfliteral{q 1 J 0.4 w 0 0 m \co@wd\space \co@ht\space l S Q}}%
  \rlap{\pdfliteral{q 1 J 0.4 w 0 \co@ht\space m \co@wd\space 0 l S Q}}%
  #1%
  \endgroup
}
\makeatother

\title{Offensive Alliances in Signed Graphs}
\date{}
\author{Zhidan Feng\inst1\inst2 \and Henning Fernau\inst2\orcidID{0000-0002-4444-3220} Kevin Mann\inst2\orcidID{0000-0002-0880-2513} \and Xingqin~Qi\inst1 } 

\institute{Shandong University, School of Mathematics and Statistic\\ 264209 Weihai, China\\
\email{qixingqin@sdu.edu.cn}\\
\and
Universit\"at Trier, Fachbereich~4 -- Abteilung Informatikwissenschaften\\  
54286 Trier, Germany.\\
\email{\{s4zhfeng,fernau,mann\}@uni-trier.de}
}

\begin{document}
\maketitle

\begin{abstract}
Signed graphs have been introduced to enrich graph structures expressing relationships between persons or general social entities, introducing edge signs to reflect the nature of the relationship, e.g., friendship or enmity. Independently, offensive alliances have been defined and studied for undirected, unsigned graphs. We join both lines of research and define offensive alliances in signed graphs, hence considering the nature of relationships. Apart from some combinatorial results, mainly on $k$-balanced and $k$-anti-balanced signed graphs (\longversion{where the latter is }a newly introduced family of signed graphs), we focus on the algorithmic complexity of finding smallest offensive alliances, looking at a number of parameterizations. While the parameter solution size leads to an \FPT result for unsigned graphs, we obtain \W{2}-completeness for the signed setting.
We introduce new parameters for signed graphs, e.g., distance to weakly balanced signed graphs, that could be of independent interest. We show that these parameters yield \FPT results. Here, we make use of the recently introduced parameter neighborhood diversity for signed graphs.
\end{abstract}

\section{Introduction}

An alliance is a grouping of entities that work together because of a mutual interest or a common purpose. The alliance might be formed to unite against potential attacks for self-defensive purposes or to establish active attacks against common enemies. An offensive alliance is a kind of alliance formed for the purpose of attacking, which has been widely studied in unsigned graphs~\cite{Favetal2002a,FerRodSig09,Har2014,Cheetal2009,HarLeg2015,Che2010,GaiMaiTri2021,CosFPRS2014}. More formally, if $G=(V,E)$ is an undirected, unsigned graph, then $S\subseteq V$ is an \emph{offensive alliance} if for each $u\in V\setminus S$, if $\deg_S(u)>0$, then $\deg_S(u)\geq \deg_{\overline S}(u)+1$, where $\deg_X(v)$ counts the number of neighbors of vertex~$v$ in the vertex set~$X$.
Here, we intend to introduce the concept of offensive alliances into signed graphs, generalizing it from the unsigned setting. 

We first list the necessary definitions. A signed graph is given by a triple $G=(V,E^+,E^-)$, where $V$ is a finite vertex set and $E^+\subseteq \binom{V}{2}$ is the positive edge set and $E^-\subseteq \binom{V}{2}$, with $E^+\cap E^-=\emptyset$, is the negative edge set. Positive edges are interpreted as friendly connections, while negative edges are rather hostile relationships.
For all $v\in V$, $N^+(v)=\{u\in V\mid vu\in E^+\}$ is the set of positive neighbors of $v$, $N^-(v)=\{u\in V\mid  vu\in E^-\}$ is the set of negative neighbors of $v$. We call the set $N(v)=N^+(v)\cup N^-(v)$ the open neighborhood of $v$, and set $N[v]=N(v)\cup\{v\}$ the closed neighborhood of $v$. Accordingly, $\deg^+_G(v)=|N^+(v)|$, $\deg^-_G(v)=|N^-(v)|$ are the positive degree of $v$ and the negative degree of $v$, respectively. Let $\deg_G(v)=\deg^+_G(v)+\deg^-_G(v)$ denote the (total) degree of~$v$. 
Let  $\delta^+(G)$ and $\Delta^+(G)$ denote the minimum and maximum positive degree of any vertex in~$G$, respectively. Accordingly, $\delta^-(G)$ and $\Delta^-(G)$ are understood, as well as $\delta(G)$ and $\Delta(G)$.
For any set $S\subseteq V$, $\deg^+_S(v)=|N^+(v)\cap S|$, $\deg^-_S(v)=|N^-(v)\cap S|$ are the positive degree and the negative degree of $v$ with respect to $S$, respectively. Let $N(S)=\bigcup_{v\in S} N(v)$ denote the neighborhood of~$S$, then the boundary of~$S$ is the set $\partial S=N(S)\setminus S$.

In this paper, we initiate the study of offensive alliances in signed graphs. Unlike unsigned setting, signed graphs distinguish the edges with the property of positive or negative signs, which models refined relationships in many real-world systems, like friendship or enmity. Consequently, this leads to different requirements on offensive alliance structures, and so we give a modified definition in the following, staying close to our proposal for a notion of defensive alliances in signed graphs, formulated in~\cite{ArrFFMQW2023}.
\longversion{\begin{definition}}
    A subset~$S\subseteq V$ of a signed graph $G=(V,E^+,E^-)$ is called an \emph{offensive alliance} if, for each $v\in \partial S$, 
    \begin{enumerate}
        \item $\deg_S^-(v)\geq \deg_S^+(v)$ and
        \item $\deg_S^-(v)\geq \deg_{\overline S}^+(v)+1$.
    \end{enumerate}
\longversion{\end{definition}}

The first condition expresses that the offensive alliance is predominantly hostile (at least not friendly) to each vertex of the boundary. It makes sure that the alliance is ready for the purpose of attacking so that friends within the alliance are staying neutral and not coming to help their friends outside the alliance. The second condition is consistent with that of a traditional offensive alliance in unsigned graphs, which says that each attacked node should have at least as many attackers in the alliance than it has potential friends outside of the alliance, which guarantees a successful attack. It can be reformulated as $\deg_S(v)\geq \deg^+(v)+1$.
A vertex is said to be successfully attacked if the vertex satisfies these two offensive conditions. We illustrate these concepts in \autoref{fig:example}. 

\begin{figure}[tbh]
    \centering
    
\begin{subfigure}[t]{.42\textwidth}
    \centering
    \begin{tikzpicture}
        \tikzset{every node/.style={fill = white,circle,minimum size=0.05cm}}

        \node[draw,label=left:$v_2$] (x2) at (0,0) {};
        \node[draw,label=left:$v_4$] (x4) at (0,-2) {};
        \node[draw,label=left:$v_1$] (x1) at (-0.9,-1) {};
        \node[draw,label=-30:$v_3$] (x3) at (0,-1) {};
        \node[draw,label=right:$v_7$] (x7) at (1,-2) {};
        \node[draw,label=160:$v_5$] (x5) at (1,0) {};

        \node[draw,label=-5:$v_6$] (x6) at (1,-1) {};

        \path (x1) edge[] (x2);
        \path (x3) edge[] (x1);
        \path (x4) edge[] (x1);
        \path (x3) edge[] (x2);
        \path (x2) edge[] (x5);
        \path (x3) edge[] (x4);
        \path (x3) edge[] (x5);
        \path (x4) edge[] (x7);
        \path (x6) edge[] (x5);
        \path (x7) edge[] (x6);

        \end{tikzpicture}
    \subcaption{Unsigned graph: $\{v_1,v_2,v_3\}$ is a minimum offensive alliance of size~$3$; it has various offensive alliances with four vertices, e.g., $\{v_1,v_3,v_5,v_7\}$,  $\{v_1,v_3,v_4,v_6\}$, $\{v_1,v_4,v_5,v_7\}$, etc., but  $\{v_1,v_3,v_4,v_5\}$ is not an offensive alliance. }  
\end{subfigure}\qquad 
\begin{subfigure}[t]{.44\textwidth}
    \centering
    \begin{tikzpicture}
        \tikzset{every node/.style={fill = white,circle,minimum size=0.05cm}}
       \node[draw,label=left:$v_2$] (x2) at (0,0) {};
        \node[draw,label=left:$v_4$] (x4) at (0,-2) {};
        \node[draw,label=left:$v_1$] (x1) at (-0.9,-1) {};
        \node[draw,label=-30:$v_3$] (x3) at (0,-1) {};
        \node[draw,label=right:$v_7$] (x7) at (1,-2) {};
        \node[draw,label=160:$v_5$] (x5) at (1,0) {};

        \node[draw,label=-5:$v_6$] (x6) at (1,-1) {};

        \path (x1) edge[dashed] (x2);
        \path (x3) edge[] (x1);
        \path (x4) edge[] (x1);
        \path (x3) edge[dashed] (x2);
        \path (x2) edge[](x5);
        \path (x3) edge[] (x4);
        \path (x3) edge[] (x5);
        \path (x4) edge[dashed] (x7);
        \path (x6) edge[dashed] (x5);
        \path (x7) edge[dashed] (x6);
        \end{tikzpicture}
    \subcaption{Signed graph: $\{v_1,v_3,v_4,v_5\}$ is a minimum offensive alliance of size~$4$, but it is not an offensive alliance in the unsigned setting; also, $\{v_1,v_2,v_3\}$ is not an offensive alliance in the signed setting, but $\{v_1,v_3,v_4,v_6\}$ still is.  }
    \label{fig:ex_signed}
\end{subfigure}
\caption{Offensive alliances on unsigned vs signed graph}
\label{fig:example}
\end{figure}

Note that a trivial case is that for each signed graph, the whole vertex set forms an offensive alliance according to the definition; more generally, a trivial alliance shall denote an alliance formed by a connected component. Unlike in the case of defensive alliance on signed graphs, the pure existence of an offensive alliance for a signed graph is therefore not a hard problem. In the following, we are going to study the following natural computational minimization problem\longversion{for offensive alliances in signed graphs}:
\begin{description}\item[\textsc{Offensive Alliance}:]
Given a signed graph $G$ and an integer $k\geq1$, does there exist an offensive alliance $S$ with $1\leq |S|\leq k$?
\end{description}

The analogously defined computational problem in unsigned graph will be called  \textsc{Offensive Alliance UG} for distinction.

\shortversion{ Due to space limitations, proofs of statements marked with $(*)$ are presented in the appendix.}

\begin{toappendix}
We now define some important (unsigned) graph parameters that we will use in this paper. To this end, let $G=(V,E)$ be an undirected unsigned graph.
\begin{description}
    \item[$\tw(G)$] A \emph{tree decomposition} of~$G$ is given by a tree $T=(V_T,E_T)$ and a mapping $f:V_T\to 2^V$ such that (1) $\bigcup_{t\in V_T}f(t)=V$, (2) for each $\{u,v\}\in E$, there is some $t\in V_T$ with $f(t)\supseteq \{u,v\}$, and (3) for each $v\in V$, $X_v=\{t\in V_T\mid f(t)\ni v\}$ is connected in~$T$. $\tw(G,T,f)=\min\{\vert f(t)\vert - 1 \mid t\in V_T\}$ is the \emph{width} of this decomposition, and the minimum width over all tree decompositions of~$G$ is the \emph{treewidth} of~$G$, written $\tw(G)$.
    \item[$\pw(G)$] If in the preceding definition, the tree happens to be a path, we speak of a path decomposition, and the minimum width over all path decompositions of~$G$ is the \emph{pathwidth} of~$G$, written $\pw(G)$.
    \item[$\td(G)$] An \emph{elimination forest} of $G$ is a rooted forest
$F$ with the same vertex set as $G$, such that for each edge $\{u,v\}$ of $G$, $u$ is an ancestor of $v$ or $v$ is an ancestor of $u$ in~$F$. (This forest can contain edges that are not in $G$.) The \emph{treedepth} of a graph $G$, denoted by $\td(G)$, is the minimum depth of an elimination forest of~$G$.
    \item[$\vc(G)$] A vertex cover of a graph $G$ is a vertex set such that each edge is incident to at least one vertex in the set. $\vc(G)$ is the minimum size of a vertex cover of $G$.    \item[$\fvs(G)$] A feedback vertex set of a graph $G$ is a vertex set such that each cycle of $G$ contains at least one vertex in the set. $\fvs(G)$ is the minimum size of a vertex cover of $G$.
\end{description}
\end{toappendix}


\section{Some Combinatorial Results}

Here, we study the size of the smallest non-empty offensive alliance in a signed graph~$G$, which we will also denote as  $a_{so}(G)$.

\longversion{\subsection{Observations}}\shortversion{\paragraph{Observations.}}
Intuitively, the more negative edges there are, the more likely it is to form the non-trivial offensive alliance. An extreme example is that the signed graph is full of positive edges, in this case, the only offensive alliance is the trivial one, \longversion{that is,} the whole vertex set. Our first observation shows a necessary relationship between negative degree and positive degree for the sake of existence of a non-trivial offensive alliance. 

\begin{observation}\label{obs:OA-degree-condition}
(necessary condition for existence) If there exists a non-trivial offensive alliance in a signed connected graph~$G$, then $\Delta^{-}(G) \geq \left\lceil \frac{\delta^{+}(G)+1}{2} \right\rceil.$
\end{observation}
\begin{pf}
    Let $S$ be a non-trivial offensive alliance of $G$. As $G$ is connected, this means that $S\neq V$. Therefore, $\partial S\neq \emptyset$. Then for any vertex $v\in\partial S$, $\Delta^{-}(G)\geq \deg^{-}({v}) \geq\deg_S^-(v)\geq \max\{\deg_S^+(v), \deg_{\overline S}^+(v)+1\}\geq \left\lceil\frac{\deg^{+}(v)+1}{2} \right\rceil\geq \left\lceil \frac{\delta^{+}(G)+1}{2} \right\rceil$.
\end{pf}

\noindent
From the previous proof, we can also deduce:

\begin{corollary}\label{cor_v}
If a vertex $v$ can be successfully attacked, (or in other words, $v$ is in the boundary of an offensive alliance), then $\deg^{-}({v}) \geq \left\lceil \frac{\deg^{+}({v})+1}{2} \right\rceil.$
\end{corollary}

\begin{proposition}\label{prop:aso-minposdeg} \shortversion{$(*)$}
    For all connected signed graphs $G$, $a_{so}(G)\geq \delta^+(G)+1$.
\end{proposition}

\begin{toappendix}
\begin{pf}\shortversion{[of \autoref{prop:aso-minposdeg}]}Let $G=(V,E^+,E^-)$ be a connected signed graph. First, assume that  $a_{so}(G)=|V|$. As (trivially) $\delta^+(G)\leq |V|-1$ (with the worst case being a positive complete graph, i.e., $E^+=\binom{V}{2}$), the inequality holds in this case. 
Assume now that a smallest offensive alliance~$S$ is not the whole vertex set. 
More generally, let $S\neq V$ be any offensive alliance of the signed connected graph~$G$. Then, $\partial S\neq\emptyset$. Hence, for each $u\in \partial S$, $\deg^-_S(u)\geq \deg^+_{\overline{S}}(u)+1$, so $|S|\geq\deg^-_S(u)+\deg^+_S(u)\geq \deg^+_{\overline{S}}(u)+1+\deg^+_S(u)=\deg^+(u)+1\geq \delta^+(G)+1$. 
\end{pf}
\end{toappendix}

Examples of equality in the previous upper bound are 2-balanced complete graphs $K_{2n}$ (see below) with two partitions of the same size, i.e.,  $|V_1|=|V_2|=n$. 

\begin{proposition}\label{prop:min-off-all-connected}
If $S$ is a minimum-size offensive alliance in~$G=(V,E^+,E^-)$, then $S\cup \partial S$ is connected in the underlying unsigned graph~$(V,E^+\cup E^-)$. 
\end{proposition}


\longversion{\subsection{Special complete signed graphs}}\shortversion{\paragraph{Special complete signed graphs.}}

A signed graph
$K=(V,E^+,E^-)$ is \emph{complete} \longversion{if $E^+$ and $E^-$ together partition $\binom{V}{2}$, i.e., }if $\binom{V}{2}=E^+\cup E^-$. Hence, for each pair of objects from~$V$, it is decided if they entertain a positive or a negative relationship. 
In the following, we determine $a_{so}$ for two special complete signed graphs:
(a) $K$ is \emph{(weakly) $k$-balanced} ($k\geq 1$) if the unsigned positive graph $K^+=(V,E^+)$ has $k$ connected components\longversion{ each of which forms a clique as $K$ is complete, so that}\shortversion{;} there is no negative edge within any such component. This concept was introduced in \cite{CarHar56,Dav67} with a sociological motivation and is also a basic notion for \textit{Correlation Clustering}~\cite{BanBluCha2004}. Quite analogously, we introduce here the following (new) class of signed graphs as follows. (b) $K$ is \emph{(weakly) $k$-anti-balanced} ($k\geq 1$) if the unsigned negative graph $K^-=(V,E^-)$ has $k$ connected components\longversion{ each of which forms a clique as $K$ is complete, so that}\shortversion{;} there is no positive edge within any such component.

\begin{toappendix}
\paragraph{A small notational comment.}
Obviously, the underlying graph of a complete  signed graph is complete. Therefore, and also to indicate its order, but slightly abusing established notation, we often write $K_n$ to refer to some complete  signed graph of order~$n$.
\end{toappendix}

\begin{theorem}\label{thm:complete-off-all}
For any signed complete graph $K_n=(V,E^+,E^-)$, $n=|V|$, we can determine its offensive alliance number in the following cases.
  \begin{itemize}
    \item [1)] If $K_n$ is (weakly) $k$-balanced ($k\geq 1$) with partition $V=(V_1,\longversion{V_2,}\dots,V_k),|V_1|\geq |V_2|\geq\dots\geq |V_k|$, then $a_{so}(K_n)=|V_1|$; $V_1$ is a minimum offensive alliance. Moreover, if $S=S_1\cup S_2\cup\dots\cup S_p, p\geq 2$, where $\emptyset\subsetneq S_1\subseteq V_{i_1}, \emptyset\subsetneq S_2\subseteq V_{i_2},\dots,\emptyset\subsetneq S_p\subseteq V_{i_p}, i_1\leq i_2\leq\dots\leq i_p$, then $|S|$ is a minimum  offensive alliance \longversion{if and only if}\shortversion{iff} $|S|=|V_1|\geq 2|S_l|$, where $l=\mathop{argmax}\limits_{j\in \{1,\longversion{2,}\dots,p\}} \{|S_j|\mid V_{i_j}\setminus S_j\neq \emptyset\}$.
    \item [2)]  If $K_n$ is (weakly) $k$-anti-balanced ($k\geq 1$) with partition $V=(V_1,\longversion{V_2,}\dots,V_k)$, and $|V_1|\geq |V_2|\geq\dots\geq |V_k|$, then we find:
    \begin{itemize}
     \item [i)] If $V=V_1\cup V_2$, and $|V_1| \geq |V_2| \geq \left\lceil\frac{n+1}{3}\right\rceil$, then $ a_{so}(K_n)=2 \left\lceil \frac{|V_1|+1}{2} \right\rceil$. Any subset $S_1$ of $V_1$ and any subset $S_2$ of $V_2$ with $|S_1|=|S_2|=\left\lceil \frac{|V_1|+1}{2} \right\rceil$, taken together, forms a minimum offensive alliance. 
     \item [ii)] If $|V_1| \geq \frac{n}{2}$, then $a_{so}(K_n)=\max\{n-|V_1|+1,2(n-|V_1|)\}$. Any subset $S_1$ of $V_1$ with $|S|=n-|V_1|$ and $V\setminus V_1$  forms a minimum offensive alliance.
     \item [iii)] Otherwise, $a_{so}(K_n)=n$.
    \end{itemize}
   \end{itemize}
\end{theorem}
\begin{pf}
    1) Let $K_n$ be $k$-balanced ($k\geq 1$), and assume that the nonempty set $S=S_1\cup S_2\cup\dots\cup S_p, p\geq 1$ is an offensive alliance of $K_n$, where $\emptyset\subsetneq S_1\subseteq V_{i_1}, \emptyset\subsetneq S_2\subseteq V_{i_2},\dots,\emptyset\subsetneq S_p\subseteq V_{i_p}, i_1\leq i_2\leq\dots\leq i_p$. If $\partial S\cap V_1=\emptyset$, clearly, $V_1\subseteq S$, so $|S|\geq |V_1|$. Otherwise, consider some arbitrary $u\in \partial S\cap V_1$.\begin{itemize}
        \item \longversion{If $i_1=1$, then}\shortversion{$i_1=1$ implies} $\deg^-_S(u)=|S|-|S_1|\geq\deg^+_{\overline{S}}(u)+1=|V_1|-|S_1|$, i.e., $|S|\geq |V_1|$.
        \item If $i_1>1$, then $\deg^-_S(u)=|S|\geq\deg^+_{\overline{S}}(u)+1=|V_1|$, i.e., $|S|\geq |V_1|$.
    \end{itemize}
     So in each case, $a_{so}(K_n)\geq |V_1|$.

    \smallskip
    \noindent        
    \underline{Case one}: Suppose $p=1$, we can easily check that each boundary vertex can be successfully attacked, then $S=V_1$ is a minimum offensive alliance.

    \smallskip
    \noindent        
    \underline{Case two}: Suppose $p\geq2$, if $S$ is a minimum offensive alliance, then $|S|=|V_1|$ (as we already know that $V_1$ is a minimum offensive alliance). For each $u\in\partial S\cap V_{i_j},j\in\{1,\dots,p\}$, $\deg^-_S(u)=|S|-|S_j|\geq\deg^+_S(u)=|S_j|$, i.e., $|S|\geq 2|S_j|$, so that $|S|=|V_1|\geq 2|S_l|$. Conversely, assume that $|S|=|V_1|\geq 2|S_l|$, where $l=\mathop{arg max}\limits_{j\in \{1,2,\dots,p\}} \{|S_j|\mid V_{i_j}\setminus S_j\neq \emptyset\}$. We can verify that~$S$ is an offensive alliance. For each $u\in\partial S\cap V_{i_j},j\in \{1,2,\dots,p\}$, $\deg^-_S(u)=|S|-|S_j|\geq |S|-|S_l|\geq|S_l|\geq |S_j|=\deg^+_S(u)$, $\deg^-_S(u)=|V_1|-|S_j|\geq |V_{i_j}|-|S_j|=\deg^+_{\overline{S}}(u)+1$.\linebreak[4] For each $u\in \partial S\cap V_h, h\notin\{i_1,\dots,i_p\}$, $\deg^-_S(u)=|V_1|\geq\max\{0,|V_h|\}=\max\{\deg^+_S(u),\deg^+_{\overline{S}}(u)+1\}$.

    \shortversion{2) For the analysis of the anti-balanced setting, we refer to the appendix.}
\begin{toappendix}
 \shortversion{\begin{pf}[of the second part of \autoref{thm:complete-off-all}] }  
    2) Assume that $K_n$ is $k$-anti-balanced ($k\geq 1$) and that the nonempty set $S=S_1\cup S_2\cup\dots\cup S_p$ is an offensive alliance of $K_n$, where $\emptyset\subsetneq S_1\subseteq V_{i_1},\dots, \emptyset\subsetneq S_p\subseteq V_{i_p}, i_1\leq \dots\leq i_p$.
    \begin{claim}
        For each $j\in\{1,\dots,k\}$, $S\cap V_j\neq \emptyset$. Therefore, $p=k$.
    \end{claim}
    \begin{pfclaim}
        Assume that $\exists j\in \{1,\dots,k\}, S\cap V_j= \emptyset$, that is, $V_j\subseteq \partial S$, then for each $u\in V_j$, $\deg^-_S(u)=0<\deg^+_{\overline{S}}(u)+1$, \longversion{leading to }a contradiction to $S$ being an\longversion{ offensive} alliance.  
    \end{pfclaim}

    For each $u\in\partial S\cap V_j$, a) $\deg^-_S(u)=|S_j|\geq\deg^+_S(u)=|S|-|S_j|$, and b) $\deg^-_S(u)=|S_j|\geq\deg^+_{\overline{S}}(u)+1=n-|S|-(|V_j|-|S_j|)+1$. So $2|V_j|\geq2|S_j|\geq^{\text{a)}}|S|\geq^{\text{b)}} n-|V_j|+1$, i.e., $|V_j|\geq\lceil\frac{n+1}{3}\rceil$. Hence, at most two different partition parts intersect the boundary of~$S$.

    \smallskip
    \noindent        
    \underline{Case one}: If $V_1\cap\partial S\neq\emptyset, V_2\cap\partial S\neq\emptyset$, we know $|V_1|\geq|V_2|\geq\lceil\frac{n+1}{3}\rceil$. From a), $|S_1|\geq \lceil\frac{|S|}{2}\rceil$ and $|S_2|\geq \lceil\frac{|S|}{2}\rceil$, so we know $S=S_1\cup S_2$ with $|S_1|=|S_2|$; implicitly, $p=k=2$. From b), $|S|\geq n-|V_2|+1=|V_1|+1$. On the other side, one can verify that any subset $S_1$ of $V_1$ and any subset $S_2$ of $V_2$ with $|S_1|=|S_2|=\lceil\frac{|V_1|+1}{2}\rceil$, taken together, forms an offensive alliance.

    \smallskip
    \noindent        
    \underline{Case two}: Assume that $V_j\subseteq\partial S$. From a), $2|S_j|\geq |S|=n-|V_j|+|S_j|$, so $|V_j|\geq|S_j|\geq n-|V_j|$, i.e., $|V_j|\geq\lceil\frac{n}{2}\rceil$; implicitly, w.l.o.g., $j=1$, so that $|S_1|\geq n-|V_1|$. Let $\overline{V_1}=V\setminus V_1$, so $|S|=|S_1|+|\overline{V_1}|\geq 2(n-|V_1|)$. Additionally, from b) $|S|\geq n-|V_1|+1$. For the special case of $n-|V_1|<1$, we have $|V_1|=n$, that is, $K_n$ is full-negative clique, so any vertex can form an offensive alliance, i.e., $a_{so}(K_n)=1=n-|V_1|+1$. Otherwise, one can verify that any subset $S_1$ of $V_1$ with $|S_1|=n-|V_1|$ and $\overline{V_1}$, taken together, forms an offensive alliance. Therefore, $a_{so}(K_n)=\max\{n-|V_1|+1,2(n-|V_1|)\}$.

    \smallskip
    \noindent        
    \underline{Case three}: If $\partial S=\emptyset$, then, in other words, $S=V$ is the only   offensive alliance. Then, $a_{so}(K_n)=n$.
\shortversion{\end{pf}}
\end{toappendix}
\end{pf}
\begin{toappendix}
It is also interesting to understand the structure of graphs when a certain (new) parameter is very small. 
\begin{theorem}\label{thm:asd=1}
 Let $G$ be a signed graph. Then,
 \begin{itemize}
  \item [1)] $a_{so}(G)=1$ \iffl $\exists v\in V(G)~\text{such that}~\forall u\in N(v), \deg^+(u)=0$.
  \item [2)] Except for case 1), $a_{so}(G)=2$ \iffl $\exists v, u\in V(G)$ such that $\forall w\in (N(v)\cup N(u))\setminus (N(u)\cap N(v)):\deg^+(w)=0$ and $\forall w\in N(v)\cap N(u):\deg^+(w)\leq 1$.
\end{itemize}
\end{theorem}
    
\end{toappendix}


\section{Classical and Fine-Grained Complexity Results}

A natural transformation is to apply known results for unsigned graphs to the case of signed graphs. Here, we first show the hardness of \textsc{Offensive Alliance} on signed graphs by a reduction from \textsc{Offensive Alliance UG}, which has been known to be \NP-complete~\cite{FerRodSig09}. This reduction allows us to further deduce parameterized hardness results with respect to a number of structural parameters.

\begin{theorem}\label{thm:off-all}
   \textsc{Offensive Alliance} 
   is \NP-complete.
\end{theorem}
\begin{pf}
For membership in \NP, a simple guess-and-check approach will work. We will prove \NP-hardness by a reduction from \textsc{Offensive Alliance UG}. 
Let $G=(V,E)$ be an unsigned graph, and $k\in\mathbb{N}$, forming an instance of \textsc{Offensive Alliance UG}. For each $v\in V$, define $d'(v)\coloneqq \left\lceil \frac{\deg(v)+1}{2}\right\rceil$, and $M_v\coloneqq \{v_{ij}\mid i\in\{1,\dots,d'(v)-1\}, j\in\{1,\dots,3k+1\}\}$. Then, in polynomial time, one can construct from $\langle G,k\rangle$ an equivalent instance $\langle G',k'\rangle$ of \textsc{Offensive Alliance}, with $G'=(V', E'^+, E'^-)$, as follows. This is also illustrated in \autoref{fig:off-all}.
    \begin{equation*}
    \begin{split}
        V' \coloneqq{}& V\cup \bigcup_{v\in V}M_v\cup \{v_1,v_2,v'_i\mid v\in V,i\in\{1,\dots,d'(v)-1\}\}\,,\\
        E'^+\coloneqq{}& \{vv’_i,v'_iv_{i1},v_{i1}v_{il},v_{i2}v_{il}\mid v\in V, i\in\{1,\dots,d'(v)-1\}, l\in\{3,\dots,3k+1\}\}\,,\\
        E'^-\coloneqq{}& E\cup \{v_1v'_i, v_2v'_i\mid v\in V, i\in \{ 1,\ldots,d'(v)-1\}\}\, ;\\
        k'\coloneqq{}& 3k\,.
    \end{split}
    \end{equation*}
%

 \begin{figure}[htp]
    \centering
\begin{subfigure}[t]{.45\textwidth}
   \centering  
   \begin{tikzpicture}[rotate=90]
        \tikzset{node/.style={fill = white,circle,minimum size=0.3cm}}

        \node[] (n1) at (0,0.7) {};
         \node[] (n2) at (0,-0.7) {};
        \node[draw,diamond] (v) at (0.5,0) {};
        \node[draw,circle] (v1) at (1,0.5) {};
        \node[draw,circle] (v2) at (1,-0.5) {};
        \node[draw,circle](z1) at (2,0.5) {};
        \node[draw,circle] (z2) at (2,-0.5) {};
        \node[draw, rectangle] (x1) at (1,1.5) {};
        \node[draw, rectangle] (x2) at (1,-1.5) {};

        \path (v) edge[dashed] (n1);
        \path (v) edge[dashed] (n2);
        \path (n1) -- node[auto=false]{\ldots} (n2);
        \path (v1) -- node[auto=false]{\ldots} (v2);
        \path (v) edge[-] (v1);
        \path (v) edge[-] (v2);
        \path (v1) edge[dashed] (z1);
        \path (v1) edge[dashed] (z2);
        \path (v2) edge[dashed] (z1);
        \path (v2) edge[dashed] (z2);
        \path (v1) edge[-] (x1);
        \path (v2) edge[-] (x2);
   \end{tikzpicture}
   \subcaption{Vertex gadget: the diamond represents vertex $v$; the square represents the not-in-the-solution gadget.}  \label{fig:OAVertexGadget}
\end{subfigure}
\begin{subfigure}[t]{.44\textwidth}
   \centering  
   \begin{tikzpicture}[rotate=270]
        \tikzset{node/.style={fill = white,circle,minimum size=0.3cm}}
        \node[draw,circle] 
        (x1) at (1,-1) {};
        \node[draw,circle] (x2) at (1,1) {};
        
        \node[draw,circle] (x3) at (0,0) {};       
        \node[draw,circle] (x4) at (0.5,0) {};
        \node[draw,circle] (x5) at (1,0) {};
        \node[draw,circle] (x6) at (2,0) {};
        \node[] (x7) at (1,-2) {}; 
        
        \path (x1) edge[-] (x3);
        \path (x1) edge[-] (x4);
        \path (x1) edge[-] (x5);
        \path (x1) edge[-] (x6);
        \path (x2) edge[-] (x3);
        \path (x2) edge[-] (x4);
        \path (x2) edge[-] (x5);
        \path (x2) edge[-] (x6);
        \path (x5) -- node[auto=false,rotate=90]{\ldots} (x6);
        \path (x1) edge[-] (x7);
   \end{tikzpicture}
   \subcaption{The not-in-the-solution gadget.}  \label{fig:OANSGadget}
\end{subfigure}    
    \caption{Reduction construction for \autoref{thm:off-all}. }
    \label{fig:off-all}
\end{figure}
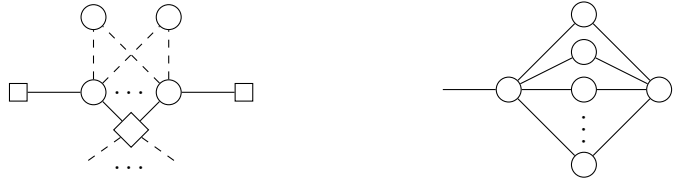

In the following, for clarity we add a subscript $G$ or $G'$ when referring to degrees, for instance.
If $S\subseteq V$ with $1\leq|S|\leq k$ is an offensive alliance of~$G$, then we prove $S'=S\cup \{v_1,v_2|v\in S\}$ is an offensive alliance of~$G'$. By definition, for each $u\in V\setminus S$ of~$G$ with $\deg_{G,S}(u)>0$, we have $\deg_{G,S}(u)\geq \deg_{G,\overline{S}}(u)+1$, that is, $\deg_{G,S}(u)\geq \left\lceil\frac{\deg_G(u)+1}{2}\right\rceil=d'(u)$. Therefore, for each $u\in \partial S$ of $G'$, if $u\in V$, then $\deg^-_{G',S}(u)=\deg_{G,S}(u)\geq d'(u)=\deg^+_{G',\overline{S}}(u)+1$. Moreover, $\deg^+_{G',S}(u)=0\leq \deg^-_{G',S}(u)=\deg_{G,S}(u)$. Otherwise, if  $u\in \partial S$ of $G'$, then $u=v'_i$ for some $v\in S$. Now, $\deg^-_{G',S}(v'_i)=2=\deg^+_{G',\overline{S}}(v'_i)+1$ and $\deg^-_{G',S}(v'_i)=2\geq\deg^+_{G',S}(v'_i)=1$. So, $S'$ is an offensive alliance on $G'$ with $1\leq|S'|\leq 3k$.

For the converse direction of this proof, \shortversion{we refer to the appendix.}\longversion{we argue as follows.}
\begin{toappendix}
\shortversion{\paragraph{Converse proof direction of \autoref{thm:off-all}.}}
Consider some vertex set $S'\subseteq V'$ with $1\leq|S'|\leq 3k$ that is an offensive alliance of $G'$. First, assume that one vertex $a$ of $M_v$ is in $S'$ for some $v\in V$. If $a=v_{il}$ for some $l\geq 3$, then $v_{i1}, v_{i2}\in S'$ as well, because if, say,  $v_{i1}\in\partial S'$, then it would have only positive neighbors in~$S'$.  If $a=v_{i1}$ (or $a=v_{i2}$), then also $v_{il}\in S'$ for any $l\geq 3$, as if  $v_{il}\in\partial S'$, then it would have only positive neighbors in~$S'$. Therefore, if $M_v\cap S'\neq \emptyset$, then 
$M_v\subseteq S'$. As $M_v$ has a size of $3k+1>k'$, then we know that for each $v\in V, M_v\nsubseteq S'$, and hence, by our previous argument, $M_v\cap S'=\emptyset$. 
Secondly, discuss
$v'_i$, where $v\in V, i\in\{1,\dots,d'(v)-1\}$. 
The vertex $v'_i$ has two positive neighbors, $v$ and $v_{i1}$, and two negative neighbors, $v_1$ and $v_2$.
As $v_{i1}\in N^+(v'_i)$ has no negative neighbors, $v'_i$ cannot be in $S'$. Hence, also none of $v'_i$ can belong to~$S'$.
Hence, if $v\in S'$, where $v\in V$, then $v'_i$ is a boundary vertex of~$S'$. As a consequence, $\{v_1,v_2\}\subseteq S'$, as otherwise the positive neighbors of $v'_i$ in $S'$ would outnumber the positive neighbors. Observe that 
if $\{v_1,v_2\}\cap S'\neq \emptyset$, then $v'_i\in\partial S'$ and $\deg^-_{G',S'}(v'_i)\geq 1$, so that  exactly one of the two positive neighbors of $v'_i$ must be in $S'$ and the other one must be in $\overline{S'}$. As we already know by our previous considerations that $v_{i1}\notin S'$, we must have $v\in S'$. Again by previous considerations, this implies $v_2\in S'$.  
That is, for each $v\in V$, $\{v,v_1,v_2\}$ must coexist in the set $S'$: either all three are in $S'$ or none of them. Then we claim that $S=\{v\in V\mid v\in S' \}$ is an offensive alliance of~$G$. For each $u$ on the boundary of~$S$, $\deg_{G,S}(u)=\deg^-_{G',S'}(u)\geq \deg^+_{G',\overline{S'}}(u)+1=d'(u)=\left\lceil \frac{\deg_G(u)+1}{2}\right\rceil$, that is, $\deg_{G,S}(u)\geq\deg_{G,\overline{S}}(u)+1$.
\end{toappendix}
\end{pf}


Next, we would like to display another simpler reduction from \textsc{Vertex Cover}, which shows the hardness of \textsc{Offensive alliance} on degree-bounded and planar signed graphs. Moreover, by this reduction, we can also get a lower-bound result on \textsc{Offensive alliance} based on the Exponential-Time Hypothesis ETH; see \cite{ImpPatZan2001,LokMarSau2011b}.
A  different reduction was used in~\cite{GaiMai2022b} for unsigned graphs.

    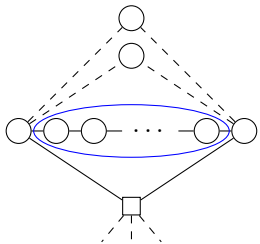
\begin{figure}[bth]
    \centering 
   \begin{tikzpicture}
        \tikzset{node/.style={fill = white,circle,minimum size=0.3cm}}

         \node[] (n1) at (-0.5,-0.6) {};
         \node[] (n2) at (0,-0.6) {};
         \node[] (n3) at (0.5,-0.6) {};
        \node[draw,rectangle] (v) at (0,0) {};

        \node[draw,circle] (i1) at (-1,1) {};
        \node[draw,circle] (i2) at (-0.5,1) {};
        \node[] (i3) at (0,1) {};
        \node[] (i4) at (0.5,1) {};
        \node[draw,circle] (i5) at (1,1) {};
        
        \node[draw,circle] (v1) at (-1.5,1) {};
        \node[draw,circle] (v2) at (1.5,1) {};
        \node[draw,circle](z1) at (0,2) {};
        \node[draw,circle] (z2) at (0,2.5) {};

        \path (v) edge[dashed] (n1);
        \path (v) edge[dashed] (n2);
        \path (v) edge[dashed] (n3);
        \path (v) edge[-] (v1);
        \path (v) edge[-] (v2);
        \path (v1) edge[-] (i1);
        \path (v2) edge[-] (i5);
        \path (i1) edge[-] (i2);
        \path (i2) edge[-] (i3);
        \path (i3) -- node[auto=false]{\ldots} (i4);
        \path (i4) edge[-] (i5);
        
        \path (v1) edge[dashed] (z1);
        \path (v1) edge[dashed] (z2);
        \path (v2) edge[dashed] (z1);
        \path (v2) edge[dashed] (z2);
        \draw[blue] (0,1) ellipse (1.3 and 0.35 );
        
   \end{tikzpicture}
    \caption{Reduction construction for \autoref{thm:OA-planar-maxdeg};  the square represents vertex $v$; the set encircled in blue is serving as a not-in-the-solution gadget as in \autoref{fig:OANSGadget}. }
    \label{fig:OA-planar-maxdeg}
\end{figure}
\begin{theorem}\label{thm:OA-planar-maxdeg}
    \textsc{Offensive Alliance} is \NP-complete, even on planar graphs with maximum degree of $5$.
\end{theorem}
\begin{pf}
\NP-membership is inherited from \autoref{thm:off-all}. We now give a simple reduction from \textsc{Vertex Cover} on planar cubic graphs (see \cite{Moh01}) to show the \NP-hardness claim.
Let $\langle G=(V,E),k\rangle$ be an instance of \textsc{Vertex Cover} on planar cubic graphs.
We can get an instance $\langle G'=(V',E'^+,E'^-),k'\rangle$, $k'= 3k$, of \textsc{Offensive alliance} in polynomial time as follows, \longversion{which is also }illustrated in \autoref{fig:OA-planar-maxdeg}:

    \begin{equation*}
    \begin{split}
        V' \coloneqq{}& V\cup \{v^1,v^2,v_1,v_2,v'_i\mid v\in V,i\in\{1,\dots,3k+1\}\}\,,\\
        E'^+\coloneqq{}& \{vv_1,vv_2,v_1v'_1,v'_iv'_{i+1},v_2v'_{4k}\mid v\in V, i\in\{1,\dots,3k\}\}\,,\\
        E'^-\coloneqq{}& E\cup \{v_1v^1, v_1v^2,v_2v^1, v_2v^2\mid v\in V\}\, ;\\
    \end{split}
    \end{equation*}
    Notice that this construction is a simplified version that of in \autoref{thm:off-all}, as the input graph $G$ with maximum degree of~$3$. Obviously, following the analogous analysis in \autoref{thm:off-all}, we can obtain $\langle G=(V,E),k\rangle$ is a  \yes-instance of \textsc{Vertex Cover} on planar cubic graphs if and only if $\langle G'=(V',E'^+,E'^-),k'\rangle$ of is a \yes-instance of \textsc{Offensive alliance}. 
In particular, if $S\subseteq V$ is a vertex cover in~$G$, then $V\cup\{v^1,v^2\mid v\in S\}$ is an offensive alliance of~$G'$. Moreover, $\Delta(G')=5$, and planarity is also shown in \autoref{fig:OA-planar-maxdeg}. 
\end{pf}

\begin{theorem}\label{thm:OA_ETH}
If ETH holds,  \textsc{Offensive Alliance} is not solvable in time $\mathcal{O}(2^{o(n)})$.
\end{theorem}
\begin{pf}
    We sightly modified the construction of \autoref{thm:OA-planar-maxdeg} by sharing a common not-in-the-solution gadget:
     \begin{equation*}
    \begin{split}
        V'_{new} \coloneqq{}& V\cup\{v^1,v^2,v_1,v_2\mid v\in V\}\cup\{c'_i\mid i\in\{1,\dots,3k+1\}\}\,,\\
        E'^+_{new}\coloneqq{}& \{vv_1,vv_2,v_1c'_1,c'_ic'_{i+1},v_2c'_{4k}\mid v\in V, i\in\{1,\dots,3k\}\}\,,\\
        E'^-_{new}\coloneqq{}& E'^-\,.
    \end{split}
    \end{equation*}
     In this way, we get an equivalent construction which has at most $5n+3k+1\leq 8n+1$ many vertices, and then we have the ETH result by~\cite{\longversion{Ami2021,}JohSze99}.
\end{pf}

\section{Parameterized Complexity and Algorithms}

It is known that \textsc{Offensive Alliance UG} is in \FPT when parameterized by solution size~\cite{FerRai07}.
The analogy to signed graphs (as indicated by the previous results) now breaks down, as we show a \W{2}-completeness result in \autoref{thm:sg-OffAll-W2}. Here, we make use of \textsc{Hitting Set}, a vertex cover problem on hypergraphs. 

\smallskip
\centerline{\fbox{\begin{minipage}{.99\textwidth}
\textbf{Problem name: }\textsc{Hitting set}\\
\textbf{Given: } A hypergraph $\mathcal{H}=(V, E)$ consisting of a vertex set $V=\{v_1,\dots, v_n\}$ and a hyperedge set $E=\{e_j\subseteq V\mid j\in \{1,\dots, m\}\}$, and $k\in \mathbb{N}$\\
\textbf{Parameter: } $k$\\
\textbf{Question: } Is there a subset $S\subseteq V$ with 
$|S|\leq k$ such that for each $e_j\in E, e_j\cap S\neq\emptyset$?\end{minipage}
}}


\smallskip
For membership, we use a reduction using the problem \textsc{Short Blind Non-Deterministic Multi-Tape Turing Machine Computation} which was introduced by Cattanéo and Perdrix in \cite{CatPer2014}. In that paper, they have also shown \W{2}-completeness of this Turing machine problem. \begin{toappendix}
The difference between a \emph{blind} multi-tape  nondeterministic  and a normal multi-tape  nondeterministic Turing machine is that the transitions can be independent of the symbols that are in the cells under the current head positions, \emph{i.e.}, the Turing machine may, but need not read the cell contents, and in this sense, it may be blind.\longversion{\footnote{Possibly, \emph{oblivious} would have been a better term for this property, but we stick to the notion \emph{blind} as introduced in the mentioned paper.}} 

\smallskip
\centerline{\fbox{\begin{minipage}{.99\textwidth}
\textbf{Problem name: }\textsc{Short Blind Non-Deterministic Multi-Tape Turing Machine Computation}\\
\textbf{Given: } A nondeterministic multi-tape Turing machine TM, a word $w$, $k\in \mathbb{N}$\\
\textbf{Parameter: } $k$\\
\textbf{Question: } Does TM accept $w$ in at most $k$ steps?\end{minipage}
}}
\end{toappendix}

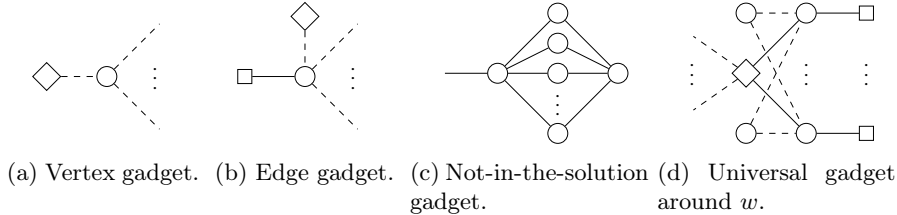
\begin{figure}[tbh]
    \centering
\begin{subfigure}[t]{.21\textwidth}
   \centering  
   \scalebox{.8}{\begin{tikzpicture}
        \tikzset{node/.style={fill = white,circle,minimum size=0.3cm}}

        \node[draw,diamond] (w) at (0,0) {};
        \node[draw,circle] (v) at (1,0) {};
        \node[] (e1) at (2,1) {};
        \node[] (e2) at (2,-1) {};

        \path (w) edge[dashed] (v);
        \path (v) edge[dashed] (e1);
        \path (v) edge[dashed] (e2);
        \path (1.8,1) -- node[auto=false,rotate=90]{\ldots} (1.8,-1);
   \end{tikzpicture}}
   \subcaption{\small Vertex gadget.}  
   \label{fig:OAVertexGadget_W2}
\end{subfigure}
\begin{subfigure}[t]{.21\textwidth}
   \centering  
    \scalebox{.8}{\begin{tikzpicture}
        \tikzset{node/.style={fill = white,circle,minimum size=0.3cm}}

        \node[draw, rectangle] (x) at (-1,0) {};
        \node[draw,circle] (e) at (0,0) {};
        \node[draw,diamond] (w) at (0,1) {};
        \node[] (v1) at (1,1) {};
        \node[] (v2) at (1,-1) {};

        \path (w) edge[dashed] (e);
        \path (e) edge[-] (x);
        \path (e) edge[dashed] (v1);
        \path (e) edge[dashed] (v2);
        \path (0.8,1) -- node[auto=false,rotate=90]{\ldots} (0.8,-1);
   \end{tikzpicture}}
   \subcaption{\small Edge gadget.}  
   \label{fig:OAEdgeGadget_W2}
\end{subfigure}
\begin{subfigure}[t]{.26\textwidth}
   \centering  
   \scalebox{.8}{\begin{tikzpicture}[rotate=270]
        \tikzset{node/.style={fill = white,circle,minimum size=0.3cm}}
        \node[draw,circle] 
        (x1) at (1,-1) {};
        \node[draw,circle] (x2) at (1,1) {};
        
        \node[draw,circle] (x3) at (0,0) {};       
        \node[draw,circle] (x4) at (0.5,0) {};
        \node[draw,circle] (x5) at (1,0) {};
        \node[draw,circle] (x6) at (2,0) {};
        \node[] (x7) at (1,-2) {}; 
        
        \path (x1) edge[-] (x3);
        \path (x1) edge[-] (x4);
        \path (x1) edge[-] (x5);
        \path (x1) edge[-] (x6);
        \path (x2) edge[-] (x3);
        \path (x2) edge[-] (x4);
        \path (x2) edge[-] (x5);
        \path (x2) edge[-] (x6);
        \path (x5) -- node[auto=false,rotate=90]{\ldots} (x6);
        \path (x1) edge[-] (x7);
   \end{tikzpicture}}
   \subcaption{\small Not-in-the-solution gadget.}  \label{fig:OANSGadget_W2}
\end{subfigure}  
\begin{subfigure}[t]{.26\textwidth}
   \centering  
    \scalebox{.8}{\begin{tikzpicture}
        \tikzset{node/.style={fill = white,circle,minimum size=0.3cm}}
       
        \node[draw,diamond] (w) at (0,0) {};
        \node[draw,circle] (p) at (0,1) {};
        \node[draw,circle] (q) at (0,-1) {};
        \node[] (v1) at (-1,0.7) {};
        \node[] (v2) at (-1,-0.7) {};
        \node[draw,circle] (u1) at (1,1) {};
        \node[draw,circle] (u2) at (1,-1) {};
        \node[draw, rectangle] (x1) at (2,1) {};
        \node[draw, rectangle] (x2) at (2,-1) {};

        \path (w) edge[dashed] (v1);
        \path (w) edge[dashed] (v2);
        \path (-0.9,1) -- node[auto=false,rotate=90]{\ldots} (-0.9,-1);
        \path (w) edge[-] (u1);
        \path (w) edge[-] (u2);
       \path (u1) -- node[auto=false,rotate=90]{\ldots} (u2);{\ldots} (0.8,-1);
        \path (x1) edge[-] (u1);
        \path (x2) edge[-] (u2);
        \path (x1) -- node[auto=false,rotate=90]{\ldots} (x2);
        \path (p) edge[dashed] (u1);
        \path (p) edge[dashed] (u2);
        \path (q) edge[dashed] (u1);
        \path (q) edge[dashed] (u2);
        
   \end{tikzpicture}}
   \subcaption{\small Universal gadget around~$w$.}  
   \label{fig:OAUniveralGadget_W2}
\end{subfigure}
    \caption{Reduction construction for \autoref{thm:sg-OffAll-W2}: the diamond represents vertex~$w$; the square represents the not-in-the-solution gadget. }
    \label{fig:OA_W2}
\end{figure}

\begin{theorem}\label{thm:sg-OffAll-W2}
   \textsc{Offensive Alliance} 
   is \W{2}-complete when parameterized by solution size.
\end{theorem}
\begin{pf}
It is well-known that \textsc{Hitting Set}, parameterized by solution size, is \W{2}-complete. We now describe a reduction that shows: $\textsc{Hitting Set}\leq_{\FPT} \textsc{Offensive Alliance}$. Let $\langle \mathcal{H}=(V,E),k\rangle$ be any instance of \textsc{Hitting Set}.  Then, in polynomial time, one can construct an equivalent instance $\langle G',k'\rangle$ of \textsc{Offensive Alliance}, with $G'=(V', E'^+, E'^-)$, as follows.
    \begin{equation*}
    \begin{split}
    M_v\coloneqq{}&  \{v^j\mid j\in\{1,\dots,5k\}\}\,,\\
        V \coloneqq{}& V\cup\{v_1\mid v\in V\}\cup\{w,p,q\}\cup \{v_e\mid e\in E\} \cup \bigcup_{e\in E}M_e\cup \bigcup_{v\in V}M_v\,,\\
        E'^+\coloneqq{}& \{wv_1, v_1v^1, v^1v^l, v^2v^l\mid v\in V, l\in \{3,\dots,5k\}\}\cup{}\\ {}& \{v_ee^1,e^1e^l,e^2e^l\mid e\in E,
        l\in \{3,\dots,5k\}\}\,,\\
        E'^-\coloneqq{}& \{wv,wv_e\mid v\in V, e\in E\}\cup\{vv_e\mid  v\in V, e\in E, v\in e\}\cup{}\\ {}&\{pv_1,qv_1\mid v\in V\}\,,\\
        k'\coloneqq{}&k+3\,.
    \end{split}
    \end{equation*}
We first show that if $\langle \mathcal{H},k\rangle$ is a \yes-instance of \textsc{Hitting Set}, then  $\langle G',k'\rangle$  is a \yes-instance of \textsc{Offensive Alliance}. If $S\subseteq V$ with $|S|\leq k$ is a hitting set of~$\mathcal{H}$, then we will prove that $S'=\{v\mid v\in S\}\cup \{w,p,q\}$ is an offensive alliance. Notice that, because $S$ is a hitting set, $$\partial S'=\{v_e\mid e\in E\}\cup\{v\mid v\in V\setminus S\}\cup \{v_1\mid v\in V\}\,.$$ 
Moreover, for each $v_e$, with $e\in E$, $w\in N^-(v_e)$ and $N^+(v_e)=\{e^1\}$, so that $\deg^-_{S'}(v_e)\geq 2=\deg^+_{\overline{S'} }(v_e)+1$ and $\deg^+_{S'}(v_e)=0$. 
For each $v$, with $v\in V\setminus S$, we have $\deg^+(v)=0$, so that $\deg^-_{S'}(v)=1= \deg^+_{\overline{S'}}(v)+1$, and $\deg^+_{S'}(v)=0$. 
Finally, for each $v_1$, with $v\in V$, $N^+(v_1)=\{w,v^1\}$ and $N^-(v_1)=\{p,q\}$, so that $\deg^-_{S'}(v_1)=2>1=\deg^+_{S'}(v_1)$, and $\deg^-_{S'}(v_1)=2=\deg^+_{\overline{S'}}(v_1)+1$. 
Hence, $S'$ is an offensive alliance of $G$ with size at most $k'=k+3$.
\smallskip

Secondly, we show that $\langle G',k'\rangle$ being a \yes-instance of \textsc{Offensive Alliance} implies that $\langle \mathcal{H},k\rangle$ is a \yes-instance of \textsc{Hitting Set}. \shortversion{Details can be found in the appendix.}
\begin{toappendix} 
\paragraph{Why $\langle G',k'\rangle$ being a \yes-instance of \textsc{Offensive Alliance} implies that $\langle \mathcal{H},k\rangle$ is a \yes-instance of \textsc{Hitting Set} in the proof of \autoref{thm:sg-OffAll-W2}.}
Assume that the signed graph~$G'$ has an offensive alliance $S'$ with $|S'|\leq k+3$. Let $e\in E$ and $v\in V$ be arbitrary.
For each $u\in M_e\cup M_v$, $\deg^-(u)=0$, while the graph induced by $M_e$ (or $M_v$) and the positive edges is connected. Hence, if one of vertices of $M_e$ or of $M_v$ in the alliance~$S'$, then all of $M_e$ (or of $M_v$, respectively) is contained in~$S'$. 
As each of these sets contains $5k>k+3$ many vertices, so $M_e\nsubseteq S'$ and $M_v\nsubseteq S'$. 
Moreover, from \autoref{cor_v}, any $u\in M_e\cup M_v$ cannot belong to the boundary of~$S'$, either, as $\deg^-(v)=0$; that is, no $v_e, e\in E$ and no $v_1, v\in V$ is contained in any offensive alliance. 
Therefore, $S'\subseteq \{v\mid v\in V\}\cup \{w,p,q\}$. 

Recall that $N^+(w)=\{v_1\mid v \in V\}$ and $N^-(w)=\{v \mid v\in V\}\cup \{v_e\mid e\in E\}$. As $N^+(w)\cap S'=\emptyset$, $\{w\}$ is not an offensive alliance.
Assume next that $w\in S'$. As $N^+_{S'}(v_1)=\{w\}$ and $N^+_{\overline{S'}}(v_1)=\{v^1\}$, $N^-(v_1)=\{p,q\}$ must be a part of $S'$. Conversely, if $\{p,q\}\cap S'\neq\emptyset$, then $u_x\in\partial S'$. As $\deg^+(v_1)=2,\deg^-(v_1)=2$, then $w$ as the only flexible positive neighbor of $v_1$ must be in $S'$. Together, this means that $\{w,p,q\}\subseteq S'$ or $\{w,p,q\}\cap S'=\emptyset$.
Assume again that $w\in S'$ and consider any $e\in E$. As $\deg^+_{\overline{S'}}(v_e)+1\geq 2$, $N^-_{S'}(v_e)$ should contain another vertex but~$w$. This could only by some $v$ with $v\in e$ by construction. Therefore, $S\coloneqq \{v\in V\mid v\in S'\}$ then forms a hitting set of the original hypergraph (encoded by~$S'$). Since $\{w,p,q\}\subseteq S'$ and $|S'|\leq k'$, $|S|\leq k'-3=k$. Conversely, if $v\in S'$, then $v_e\in\partial S'$ for those hyperedges $e$ that contain~$v$. As $N^+(v_e)=\{e^1\}$, a further vertex from $\{w\}\cup\{v\mid v'\in e\setminus\{v\}\}$ must be in $S'$. However, if $w\notin S'$, then $w\in\partial S'$. As $\deg_{\overline{S'}}^+(w)=|X|$, we should find $\deg_{S'}^-(w)>|X|$, which is impossible. Hence, $w\in S'$ is enforced by $v\in S'$. Therefore, any non-empty offensive alliance $S'$ must contain $\{w,p,q\}$ and an encoding of a hitting set of the given hypergraph as described above.
\end{toappendix}
\longversion{\smallskip}

\shortversion{The proof of membership in \W{2} can be also found in the appendix.}
\begin{toappendix}
\shortversion{\paragraph{Supplementing the proof of \autoref{thm:sg-OffAll-W2}}}
Next, we prove membership in \W{2}. We do this by describing how to build a \textsc{Short Blind Non-Deterministic Multi-Tape Turing Machine Computation} instance $\langle M,\varepsilon,k'\rangle$, based on the instance $\langle G,k\rangle$ of \textsc{Offensive Alliance}, where $G=(V,E^+,E^-)$.
The Turing machine $M$ should work as follows (and this has to be ensured by the reduction):
\begin{itemize}
    \item $M$ has $2\cdot \vert V\vert+1$ many tapes. We call them $t_0$, $t_v$ and $t_v'$ (for $v\in V$ in the following. 
    \item On $t_0$, $M$ works with the tape alphabet~$V$, while on the other tapes, it uses the tape alphabet $0,1$.
    \item Initially, in preparation phase~1, $M$ (blindly) writes $0^k1^{k+1}$ on each of the tapes $t_v$ and moves the head back on the first occurrence of $1$; this can be done in $3k+1$ steps by using parallelism.
    \item Then, in preparation phase~2, $M$ (blindly) writes $\min \{\deg^+(v)+1,k+2\}$ many zeros, followed by $2k+4-\min \{\deg^+(v)+1,k+2\}$ many ones, on tape~$t_v'$. Then, $M$ moves back onto the first symbol of that tape. The reduction machine will construct $M$ in a way that  all movements are done in parallel for all tapes $t'_v$, so that $M$ will need $4k+8$ many steps. This is possible as the reduction machine can compute the quantity $\min \{\deg^+(v)+1,k+2\}$ for each tape (in polynomial time) by analyzing $\langle G,k\rangle$.
    \item Now, $M$ will guess at most $k$ many  different symbols from~$V$ and write them on tape $t_0$. Clearly, this guessing takes at most $k$ steps. The guessed symbols should comprise the offensive alliance $S$. The reduction machine has to make sure by its construction that the guessed symbols are pairwisely different. The alliance property (i.e., the offensive conditions) still needs to be checked in what follows.
    \item To this end, $M$ then reads these symbols sequentially. Upon reading symbol $v\in S$, $M$ will do the following (again using parallelism, assuming a careful construction by the reduction machine):
    \begin{itemize}
        \item For all $u\in N^+(v)$ (which clearly are in $\partial S$), $M$ (blindly) moves the head on $t_u$ one step to the left and the head on $t_u'$ one step to the right.
        \item For all $u\in N^-(v)$ (which clearly are in $\partial S$), $M$ (blindly) moves the head on $t_u$ and on $t'_u$ one step to the right.
        \item On tape $t_v$ and on tape $t_v'$, $M$ writes $2k+1$ many ones and then moves $k$ steps left.
    \end{itemize}
    \item Finally, i.e., after $k$ parallel steps (as described), $M$ will accept if and only if each of the heads on $t_x$ and $t_x'$ (for each $x\in X$) reads the symbol $1$. By double-counting, one can see that, for each $u\in V$, in the end the head on tape $t_u$ moved $\deg_S^-(u)-\deg_S^+(u)$ many steps to the right (where negative steps to the right are steps to the left), compared to their position after preparation phase~1, while the head on tape $t'_u$ moved $\deg_S(u)$ many steps to the right, compared to their position after preparation phase~2. Hence, only if $\deg_S^-(u)-\deg_S^+(u)\geq 0$ and if $\deg_S(u)\geq \min \{\deg^+(v)+1,k+2\}$, $M$ will accept. Notice that since $S$ has at most $k$ elements, this last test is equivalent to testing to   $\deg_S(u)\geq \deg^+(v)+1$. So, all conditions are correctly tested if $u\in\partial S$. If $u\in S$, then we have made sure that tape $t_u$ contains sufficiently many ones so that it will end up on a one even though after reading~$u$, the heads on $t_u$ and on $t_u'$ can still move due to reading vertices from $N_S(u)$. Finally, if $u\in(\overline{S}\setminus \partial S)$, then the heads will not move at all.
\end{itemize}
Altogether, one can define some $k'\in\mathcal{O}(k^2)$ such that $M$ accepts within $k'$ steps if and only if  $G=(V,E^+,E^-)$ possesses an offensive alliance of size at most~$k$.
\end{toappendix}
\end{pf}

Based on structural hardness results for \textsc{Offensive Alliance UG} from~\cite{GaiMai2021}, the reduction presented in \autoref{thm:off-all} (or a slight variation thereof) shows:
\begin{corollary}\label{cor:OffAllStructural}\shortversion{$(*)$}
    \textsc{Offensive Alliance} on signed graphs is \W{1}-hard when parameterized by any of the following parameters\shortversion{, defined in the appendix}:
    \begin{itemize}
        \item feedback vertex set number of the underlying unsigned graph,
        \item treewidth and pathwidth of the underlying  unsigned graph,
        \item treedepth of the underlying   unsigned graph.
    \end{itemize}
\end{corollary}

\begin{toappendix}
\begin{pf}\shortversion{[of~\autoref{cor:OffAllStructural}]}
    We get this corollary from~\cite{GaiMai2021,GaiMai2022b}, which shows that \textsc{Offensive Alliance UG} is W[1]-hard parameterized by a wide range of structural parameters: the feedback vertex set number, treewidth, pathwidth, and treedepth of the input graph. Next, we will show that all these parameters of the underlying unsigned graph of our constructed signed graph in \autoref{thm:off-all} are all bounded by simple functions of the same parameters of the unsigned graph that we started with. Let $\fvs(G),\tw(G),\pw(G),\td(G)$ denote the feedback vertex set number, treewidth, pathwidth, and treedepth of the unsigned graph~$G$, respectively. 
    As in the proof of the preceding theorem, let $G=(V,E)$ be an unsigned graph and let $G'$ be the signed graph constructed from $G$.
    
    We first show that the pathwidth of the underlying graph of $G'$, $\pw(G')$, depends only on $\pw(G)$. We do so by modifying an optimal path decomposition $(\mathcal{P},\{X_p\}^n_{p=1})$ of $G$: for each $v\in V$, we take an arbitrary bag $X_j\ni v$, and add a chain of bags $X_j,X_j\cup \{v_1,v_2,v'_i,v_{i1}\}, X_j\cup \{v_1,v_2,v_{i1},v_{i2},v_{i3}\}, X_j\cup \{v_1,v_2,v_{i1},v_{i2},v_{i4}\},\dots,X_j\cup \{v_1,v_2,v_{i1},v_{i2},v_{i4k}\},X_j$ after a bag $X_j$, successively according to the order of $i\in \{1,\dots,d'(v)\}$. It is easy to verify the result is a valid path decomposition of the underlying graph of $G'$, so $\pw(G')\leq \pw(G)+5$. As path decomposition is a special tree decomposition. Hence similarly, we can obtain $\tw(G')\leq \tw(G)+5$. 

    Then we prove that $\td(G')\leq \td(G)+6$. Let $T_G=(V,E)$ be an elimination tree with the minimum depth of~$G$. We modify $T_G$ into an elimination tree of~$G'$ as follows: for each $v\in V$, take $v$ as a parent-root node and add a child-subtree $T_v=(V_v,E_v)$, where $V_v=M_v\cup\{v_1,v_2,v'_i\mid i\in\{1,\dots,d'(v)-1\}\}$, $E_v=\{vv_1,v_1v_2,v_2v'_i,v'_iv_{i1},v_{i1}v_{i2},v_{i2}v_{il}\mid i\in\{1,\dots,d'(v)-1\},l\in\{3,\dots,4k\}\}$, as illustrated in \autoref{fig:par-treedepth}. Obviously, $\td(G')\leq \td(G)+6$. 

    \begin{figure}[htp]
    \centering 
   \begin{tikzpicture}
        \tikzset{node/.style={fill = white,circle,minimum size=0.3cm}}

        \node[] (v0) at (-1,0) {};
        \node[draw, rectangle] (v) at (0,0) {};
        \node[draw,circle] (v1) at (1,0) {};
        \node[draw,circle] (v2) at (2,0) {};
        \node[draw,circle] (vi1) at (3,1) {};
        \node[draw,circle] (vi2) at (3,-1) {};
        \node[draw,circle] (vi11) at (4,1) {};
        \node[draw,circle] (vi21) at (4,-1) {};
        \node[draw,circle] (vi12) at (5,1) {};
        \node[draw,circle] (vi22) at (5,-1) {};
        \node[draw,circle] (vi13) at (6,1.5) {};
        \node[draw,circle] (vi14) at (6,0.5) {};
        \node[draw,circle] (vi23) at (6,-0.5) {};
        \node[draw,circle] (vi24) at (6,-1.5) {};

        \path (v) edge[-] (v0);
        \path (v) edge[-] (v1);
        \path (v2) edge[-] (v1);
        \path (v2) edge[-] (vi1);
        \path (v2) edge[-] (vi2);
        \path (vi1) edge[-] (vi11);
        \path (vi2) edge[-] (vi21);
        \path (vi11) edge[-] (vi12);
        \path (vi21) edge[-] (vi22);
        \path (vi12) edge[-] (vi13);
        \path (vi12) edge[-] (vi14);
        \path (vi22) edge[-] (vi23);
        \path (vi22) edge[-] (vi24);
        \path (vi2) -- node[auto=false,rotate=90]{\ldots} (vi1);
        \path (vi13) -- node[auto=false,rotate=90]{\ldots} (vi14);
        \path (vi23) -- node[auto=false,rotate=90]{\ldots} (vi24);      
   \end{tikzpicture}
    \caption{A child-subtree with parent $v$, where the square represents vertex $v$. }
    \label{fig:par-treedepth}
\end{figure}
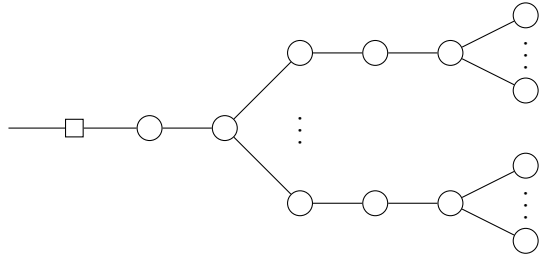

    As for the feedback vertex set number, we need to sightly modify the reduction by sharing a common not-in-the-solution gadget $M=\{v_{1j}|j\in\{1,\dots,4k\}\}$ and $v_1,v_2$ for each $v\in V$, (which make no difference on original reduction proof), producing the signed graph $G'_{\text{fvs}}$. Then clearly, deleting three more vertices $\{v_1,v_2,v_{11}\}$ leads to the underlying graph without cycles, that is, $\fvs(G'_{fvs})\leq \fvs(G)+3$. 
\end{pf}
\end{toappendix}

As the preceding result implies that \textsc{Offensive Alliance} problem parameterized by treewidth of the input graph is a hard problem, we look at a bigger parameter called domino treewidth. We will show that when parameterized by domino treewidth~$d$, with a suitably given decomposition, the problem of finding smallest offensive alliance on signed graphs is fixed parameter tractable. 

\longversion{\begin{definition}}
 A tree decomposition $(T, \{X_t\}_{t\in V_T})$ is a \emph{domino tree decomposition} if, for $i, j \in V_T$, where $i \neq j$ and $\{i, j\} \notin E_T$, we find $X_i\cap X_j = \emptyset$.
\longversion{\end{definition}} 
In other words, in a domino tree decomposition, every vertex of $G$ appears in at most two bags in~$T$.
The \emph{domino treewidth} $\dtw(G)$ is defined as the minimum \emph{width} $\max_{t\in V(T)}|X_t|-1$ over all domino tree decompositions\longversion{ $(T, \{X_t\}_{t\in V_T})$}.

\begin{theorem}\label{thm:dtw}
 \textsc{Offensive Alliance} on signed graphs is \FPT\longversion{ when}\shortversion{,} parameterized by the width of a given domino tree decomposition of the underlying unsigned graph.
\end{theorem}


\begin{pf}
Let $G=(V, E^+, E^-)$ be an instance of \textsc{Offensive Alliance} 
and $(T, \{X_t\}_{t\in V_T})$ be a domino tree decomposition of $G_0=(V, E^+\cup E^-)$ of width~$d$. 
We can assume that $G$ is connected, as otherwise we can compute $a_{so}(G)$ by considering the connected components separately.  Suppose $T=(V_T,E_T)$ is rooted at node $r$. For a node $t\in V_T$, let $T_t$ be the subtree of $T$ rooted at $t$, with vertex set $V(T_t)$, and $V_t=\bigcup_{t'\in V(T_t)}X_{t'}$ be the union of all bags of the subtree. We denote by $p(t)$ the parent node of~$t$.

Let $X_t$ be a non-leaf bag, then a vertex $v\in X_t$ can be of three types.\\ \emph{Type 1}: $v$ is also in the bag of one of the children of $t$;\\ \emph{Type 2}: $v$ is also in the bag of the parent of $t$;\\ \emph{Type 3}: $v$ is only in $X_t$.

The $\FPT$ algorithm will be a dynamic program. In this algorithm, we will inductively compute the values $c[t,A_t]$ for $t\in V_T$ and $A_t\subseteq X_t$. For the values of $c[t,A_t]$, we differentiate if $A_t=\emptyset$ or not in the following semantic definition. 

If $A_t$ is empty, $c[t,\emptyset]$ denotes the minimum cardinality of an offensive alliance $S_t$ with $S_t\subseteq V_t\setminus X_{p(t)}$.  
    
If $ \emptyset \neq A_t \subseteq X_t$, we need\longversion{ to introduce some} more notation: we record a potential offensive alliance (pOA) on the subtree $T_t$, which is a smallest non-empty set 
$S_t\subseteq V_t$ with $S_t\cap X_t=A_t$, and for each $u\in \partial S_t\setminus X_{p(t)}$: 1) $\deg_{S_t}^-(u)\geq \deg_{S_t}^+(u)$; 2) $\deg_{S_t}^-(u)\geq \deg_{\overline {S_t}}^+(u)+1$. That is, the pOA can successfully attack the boundary vertices of $S_t$ except for those falling in the class of Type~$2$ with regard to $X_t$. Let $c[t, A_t]=|S_t|$ denote the size of $S_t$; if no such set exists, we set $c[t,A_t]=\infty$.

We can use bottom-up dynamic programming to obtain these table values\longversion{, as follows}: 

    \begin{itemize}
        \item If $t$ is a \underline{leaf node}, then for each  $A_t \subseteq X_t$, $A_t\neq\emptyset$, 
        we define $c[t, A_t]$ as follows:      
        $$ c[t,A_t]\coloneqq \begin{cases}
    |A_t|, & \text{if }A_t~\text{can successfully attack all vertices in}~\partial A_t\setminus X_{p(t)} \\
    \infty, & \text{otherwise}\\
\end{cases}$$

        \item  If $t$ is an \underline{inner node} with children $t_1,t_2,\dots,t_m$, for each $A_t\subseteq X_t$, $A_t\neq\emptyset$,  let $A_{t_j}\subseteq X_{t_j}, j\in\{1,\dots,m\}$, then $A_t$ is said to be compatible with $A_{t_1},A_{t_2},\dots,A_{t_m}$ if and only if:
        \begin{itemize}
            \item [1)] For $1\leq j\leq m$, $A_t\cap X_t\cap X_{t_j}=A_{t_j}\cap X_t\cap X_{t_j}$;
            \item [2)] Type~$1$ and Type~$3$ vertices of $\partial S_t\cap X_t$ are successfully attacked by $A_t\cup A_{t_1}\cup\dots\cup A_{t_m}$.
        \end{itemize}
        For $A_t\subseteq X_t$, $A_t\neq\emptyset$, if there does not exist any $A_{t_j}\subseteq X_{t_j}$ that is compatible with $A_t$, then $c[t, A_t] = \infty$. Otherwise,
        \begin{equation}\label{eq:dtw-recursion}
            \begin{split}
            {}&c[t,A_t]\coloneqq  |A_t|+\min\{{} \sum^m_{j=1,A_{t_j}\neq\emptyset}(c[t_j, A_{t_j}]-|A_t\cap A_{t_j}|)\mid{} \\ {}& \forall 1\leq j\leq m, A_{t_j}\subseteq X_{t_j}  \implies A_t ~\text{is compatible with}~A_{t_1},\dots ,A_{t_m} \}.
            \end{split}
        \end{equation}
    
     \end{itemize}
    Similarly the values of  $c[t,\emptyset]$ can be computed. \shortversion{For details, see the appendix.}
    \begin{toappendix}
    \shortversion{\paragraph{Computation of $c[t,\emptyset]$ in the proof of \autoref{thm:dtw}}}
         \begin{itemize}
        \item If $t$ is a \underline{leaf node}, we define $c[t, \emptyset]$ as follows:      
        $$ c[t,\emptyset]\coloneqq \begin{cases}
        \begin{split}
            \min\{|B_t|\mid{}&  B_t \subseteq X_t\setminus X_{p(t)}, B_t\neq \emptyset, \\{}& B_t~\text{can successfully attack all vertices in}~\partial B_t\}
        \end{split}
     \\
    \infty,  \quad \text{otherwise}\\
\end{cases}$$

        \item  If $t$ is an \underline{inner node} with children $t_1,\longversion{t_2,}\dots,t_m$, for each $B_t\subseteq X_t\setminus X_{p(t)},B_t\neq\emptyset$, let $B_{t_j}\subseteq X_{t_j},j\in\{1,\dots,m\}$, then $B_t$ is called formative with $B_{t_1},\longversion{B_{t_2},}\dots,B_{t_m}$ \iffl
        \begin{itemize}
            \item [1)] For $1\leq j\leq m$, $B_t\cap X_t\cap X_{t_j}=B_{t_j}\cap X_t\cap X_{t_j}$;
            \item [2)] $B_t\cup B_{t_1}\cup\dots\cup B_{t_m}$ can successfully attack all the vertices in $\partial S_t\cap X_t$.
        \end{itemize}
        Let       
        \begin{equation}\label{eq:dtw-emptyset_b}
            \begin{split}
             b_t{}&\coloneqq  |B_t|+\min\{\sum^m_{j=1,B_{t_j}\neq \emptyset}(c[t_j, B_{t_j}]-|B_t\cap B_{t_j}|)\mid \\ {}&\forall 1\leq j\leq m, B_{t_j}\subseteq X_{t_j}  \implies B_t ~\text{is formative with}~B_{t_1},\dots ,B_{t_m} \}.
            \end{split}
        \end{equation}
        %
then
        \begin{equation}\label{eq:dtw-emptyset_c}
            c[t,\emptyset]=\min_{j\in\{1,\dots,m\}}\{c[t_j,\emptyset],b_t
            \}.
        \end{equation}      
     \end{itemize}
    \end{toappendix}
     \begin{claim}\label{thm:dtw_claim}\shortversion{$(*)$}
         For every node $t$ in $T$ and every $ A_t\subseteq X_t$, with $A_t\neq\emptyset$, $c[t, A_t]$ is the size of the smallest potential non-empty offensive alliance $S_t$ where $S_t\subseteq V_t$ and $S_t \cap X_t=A_t$, while $c[t,\emptyset]$ is the size of smallest offensive alliance $S_t$ where $S_t\subseteq V_t\setminus X_{p(t)}$ if it exists. Further, the size of the minimum non-empty offensive alliance in $G$ is $c[r, \emptyset]$, where $r$ is the root node of $T$. 
     \end{claim}
     
     \begin{toappendix}
     \begin{pfclaim}\shortversion{[of the Claim in \autoref{thm:dtw}]} 
         We give the proof by using  mathematical induction on the nodes of $T$. 
         \begin{itemize}
             \item For $t\in V_T$, where $t$ is a leaf node, for $A_t\subseteq X_t$, we have two subcases.
             \begin{itemize}
             \item If $A_t\neq\emptyset$, $A_t$ is compatible with the empty set, then $c[t, A_t]=|A_t|$; otherwise, $c[t,A_t]=\infty$.

             \item If $A_t=\emptyset$, if there exists non-empty $B_t\subseteq X_t\setminus X_{p(t)}$ which can successfully attack all boundary vertices, then $c[t,\emptyset]$ equals to the minimum size of such $B_t$; otherwise, $c[t,\emptyset]=\infty$.
         \end{itemize}
           Thus the base case holds.
            
            \item Let $T_t$ be a subtree rooted at an inner node $t$ with children $t_1,t_2,\dots,t_m$. Let $T_{t_1}, T_{t_2},\dots,T_{t_m}$ be the subtrees rooted at $t_1,t_2,\dots,t_m$, respectively. Assume for $1\leq j\leq m$, the statement holds for $t_j$, which means two different cases.
            \begin{itemize}
                \item For all $1\leq j\leq m$, where $A_{t_j}\subseteq X_{t_j}$, $c[t_j, A_{t_j}]$ is the size of the smallest potential offensive alliance $S_{t_j}\subseteq V_{t_j}$ where $S_{t_j}\cap X_{t_j}=A_{t_j}$. We then show that for every $A_t\subseteq X_t$, $c[t,A_t]$ is the size of the smallest potential offensive alliance $S_t$ of $T_t$ where $S_t \cap X_t=A_t$. If subsets of $X_{t_1},\dots, X_{t_m}$ cannot be found such that $A_t$ is compatible with them, then $c[t, A_t]=\infty$. Otherwise, the recurrence formula from \autoref{eq:dtw-recursion} applies.  
                

        \hspace{0.4cm} To prove that $S_t$ is a smallest pOA, we need to show that each $u\in \partial S_t\setminus X_{p(t)}$ can be successfully attacked. For $1\leq j\leq m$, let $A_{t_j}\subseteq X_{t_j}$ be compatible with~$A_t$. Thus, all the boundary  vertices of $S_t$ falling in the classes of Type~$1$ and Type~$3$ in $X_t$ satisfy both offensive conditions. Moreover, for $1\leq j\leq m, A_{t_j}\subseteq X_{t_j}$ was selected such that $c[t_j, A_{t_j}]$ is minimum over all subsets of $X_{t_j}$. From the inductive hypothesis, for $1\leq j\leq m$, the boundary vertices of $S_{t_j}$ except for those falling in the class of Type~$2$ could be successfully attacked. By definition, the vertices of Type~$2$ in $X_{t_j}$ are simultaneously the vertices of Type~$1$ in $X_t$. Thus,
        all $u\in \partial S_t\setminus X_{p(t)}$ could be successfully attacked. Furthermore, $c[t_j, A_{t_j}]$ is the size of a smallest pOA $S_{t_j}$ in $V_{t_j}$, where $S_{t_j}\cap X_{t_j}=A_{t_j}$. Therefore, $c[t, A_t]$ is the minimum size of a pOA in $G_t$, where $S_t\cap X_t=A_t$.
        \item For all $1\leq j\leq m$, $c[t_j,\emptyset]$ is the minimum size of non-empty offensive alliance lying in $V_t\setminus X_t$. Then for $t$, if there does not exist offensive alliances, (implying all $c[t_j,\emptyset]=\infty$ currently), then $c[t,\emptyset]=\infty$; otherwise we can get the number by \autoref{eq:dtw-emptyset_b},\autoref{eq:dtw-emptyset_c}.

        \hspace{0.4cm} We consider the existence of offensive alliances in the following cases:

        \underline{Case one:} If there is an offensive alliance with vertices coming from $X_t\setminus X_{p(t)}$, we need to show that each $u\in\partial S_t$ can be successfully attacked. As $B_t ~\text{is formative with}~B_{t_1},\dots ,B_{t_m}$, if some $B_{t_j}\neq\emptyset$, we know all the boundary vertices of $S_t$ in $V_{t_j}\setminus X_t$ are successfully attacked (from compatiblity). Then from formativeness, we know that all the boundary vertices in $X_t$ are successfully attacked. Base on \autoref{eq:dtw-emptyset_b}, $b_t$ is the smallest size of these offensive alliances. 
        
        
        \underline{Case two:} If there is an offensive alliance lying in $V_{t_j}\setminus X_{t},\forall j\in\{1,\dots,m\}$, it would also be an offensive alliance lying in $V_t\setminus X_{p(t)}$. 
        
        \hspace{0.4cm} In conclusion, we record a minimum one among all these possibilities.
            \end{itemize}
         \end{itemize}      
        
        Since $r$ does not possess a parent node, $c[r,\emptyset]$ is the minimum size of an offensive alliance in the whole signed graph~$G$.
     \end{pfclaim}
     \end{toappendix}
     
     Note that at an inner node, we compute (at most) $2^{d+1}$ many $c[\cdot,\cdot]$ values and the time needed to compute each of these values is $\mathcal{O}(2^{m(d+1)}\cdot d^2)$. The number of nodes in a domino tree decomposition is $\mathcal{O}(n)$. Also, $m\leq d$, because each child node has disjoint bags but must have common vertices with the bag of its parent since we assume that the signed graph~$G$ is connected. Hence, the total running time of the algorithm is $\mathcal{O}(d^22^{\mathcal{O}(d^2)}n)$.
\end{pf}

In ~\cite{ArrFFMQW2023}, when studying the defensive alliance theory on signed graphs, a new structural parameter for signed graphs was introduced:  
\longversion{\begin{definition}}
    The \emph{signed neighborhood diversity} of a graph $G = (V,E^+,E^-)$, denoted by $\snd(G)$, is the least integer $k$ for which we can partition the set $V$ of vertices into $k$ classes, such that for each class $C_i$,  any pair of vertices $\{u,v\}\subseteq C_i$ satisfies $N^+(v)\setminus\{u\}=N^+(u)\setminus\{v\}$ and $N^-(v)\setminus\{u\}=N^-(u)\setminus\{v\}$.
\longversion{\end{definition}}
If the signed neighborhood diversity of a signed graph is $k$, there is a partition $\{C_1, C_2,...,C_k\}$ of $V$ into $k$ equivalence classes. Notice that each class could be a positive clique, a negative clique or an independent set by definition. Moreover, it is not difficult to see that, between any two different classes $C_i,C_j$, the potential edges $\{u,v\}$ with $u\in C_i$ and $v\in C_j$ either all belong to $E^+$, or all belong to $E^-$, or all belong to $\binom{V}{2}\setminus \left(E^+\cup E^-\right)$. 
Here, we will show that \textsc{Offensive Alliance} on signed graphs is also \FPT when parameterized by signed neighborhood diversity. The idea is to formulate an equivalent Integer Linear Program (ILP), with a linear number of variables in this parameter, and to then resort to \cite[Theorem 6.5]{CygFKLMPPS2015}.



\begin{theorem} \shortversion{$(*)$} \label{thm:OffAllSND}
    \textsc{Offensive Alliance} 
    is \FPT when parameterized by the signed neighborhood diversity.
\end{theorem}
\begin{toappendix}
\begin{pf}\shortversion{[of \autoref{thm:OffAllSND}]}
Let $G=(V,E^+,E^-)$ be a signed graph with $k=\snd(G)$. 
We now construct an ILP with $3k$ many variables to find a smallest non-empty offensive alliance $S$ of~$G$. Our \FPT result comes from employing \cite[Theorem 6.5]{CygFKLMPPS2015}, which shows that one can solve ILP in  \FPT-time with the number of variables as the parameter.  
    
    Let $ \{1,\dots,k\}=:\mathcal{K}=\mathcal{P}\cup\mathcal{N}\cup\mathcal{I}$, where \begin{eqnarray*}
    \mathcal{P}&=&\{j\in\mathcal{K}\mid C_j ~\text{is a positive clique}\}\,,\\
    \mathcal{N}&=&\{j\in\mathcal{K}\mid C_j ~\text{is a negative clique}\}\,, \text{and}\\
    \mathcal{I}&=&\{j\in\mathcal{K}\mid C_j ~\text{is an independent set}\}\,.
    \end{eqnarray*} Furthermore, for each $i\in\mathcal{K}$, we keep a Boolean constant~$z_i$, where  $$z_i=\left\{\begin{array}{ll}
         0,&  \text{ if } i\in \mathcal{N}\cup\mathcal{I}\,,\\
         1,& \text{ if } i\in\mathcal{P}\,.
    \end{array}\right.$$ For each $i\in \mathcal{K}$, define $N^+_i\coloneqq\{j\in\mathcal{K}\mid C_j\subseteq N^+(C_i)\}$ and $N^-_i\coloneqq\{j\in\mathcal{K}\mid C_j\subseteq N^-(C_i)\}$.  Then we associate a variable $x_i\in\{1,\dots,|C_i|\}$ for each $i\in\mathcal{K}$, which indicates $|C_i\cap S|=x_i$. We also introduce two additional variables $w_i\in\{0,1\}$, and $y_i\in\{0,1\}$ for each $i\in\mathcal{K}$, wherein $w_i$ represents whether there exist vertices of $C_i$ in $\partial S$, except for the case of $C_i\subseteq S$, this would be explained in $y_i$, that is, $y_i=1$ if $x_i=|C_i|$, otherwise $y_i=0$. 
    
    Then our ILP formulation of \textsc{Offensive Alliance} is given by:

    \begin{align}
        \min  & \sum^k_{i=1} x_i \label{snd_ob}\\
        \text{s.t.} & \sum_{j\in N^-_i} x_j\geq\sum_{j\in N^+_i}x_j-2(1-w_i+y_i)|V|& \forall i\in\mathcal{K} \label{snd_inn_cond}\\
        {}& \sum_{j\in N^-_i} x_j\geq\sum_{j\in N^+_i}(|C_j|-x_j)+1-z_i-2(1-w_i+y_i)|V| & \forall i\in\mathcal{K}\label{snd_ext_cond}\\
        {}& \sum\limits_{j\in N^+_i\cup N^-_i\cup \{i\}} x_j\leq |V| w_i\leq |V|\sum\limits_{j\in N^+_i\cup N^-_i\cup \{i\}} x_j & \forall i\in\mathcal{P}\cup\mathcal{N}\label{snd_bound}\\
        {}& \sum\limits_{j\in N^+_i\cup N^-_i} x_j\leq |V|w_i\leq 
 |V|\sum\limits_{j\in N^+_i\cup N^-_i} x_j & \forall i\in\mathcal{I}\label{snd_bound_I}\\
        {}& x_i-|C_i|+1\leq y_i\leq \frac{x_i}{|C_i|} & \forall i\in\mathcal{K}\label{snd_total}\\
        {}& \sum^k_{i=1}x_i\geq 1\label{snd_nonemty}\\
        {}& x_i\in\{0,1,\dots,|C_i|\},w_i,y_i\in\{0,1\} & \forall i\in\mathcal{K}
    \end{align}
    
    For $i\in\mathcal{P}\cup\mathcal{N}$, $C_i\cap\partial S\neq \emptyset$ \iffl the closed neighborhood of $C_i$, when $C_i$ is a positive or negative clique, provides vertices to the solution, this is maintained in \autoref{snd_bound}, where for $w_i\in\{0,1\}$, $w_i=0$ \iffl $\sum_{j\in N^+_i\cup N^-_i\cup \{i\}} x_j=0$ (analogous to the case of $i\in\mathcal{I}$ shown in \autoref{snd_bound_I}). Moreover, we should pay attention to the case of $C_i\subseteq S$, which also means there is no boundary vertices in $ C_i$. Then \autoref{snd_total} guarantees that $y_i=1$ \iffl $x_i=|C_i|$; otherwise, $y_i=0$. Then we could verify the two offensive conditions (i.e., \autoref{snd_inn_cond},\autoref{snd_ext_cond}) for the boundary vertices. For each $i\in\mathcal{K}$, if $w_i=0$ or $y_i=1$, from the above, we know that this means there is no boundary vertices in $C_i$, so we need not check the offensive conditions for~$i$, and at this time, \autoref{snd_inn_cond} and \autoref{snd_ext_cond} always hold; otherwise, $w_i=1$ and $y_i=0$, so that $2(1-w_i+y_i)|V|$ vanishes. Then,  for each $i\in\mathcal{K}$, there exists a boundary vertex $v\in C_i\cap \partial S$. If $C_i\subseteq N^-_i$,  $\deg^-_{S\cap C_i}(v)=x_i$. Moreover, for $j\in N^-_i\setminus\{i\}$, $\deg^-_S(v)=x_j$, so $\deg^-_S(v)=\sum_{j\in N^-_i} x_j$; similarly, we know $\deg^-_S(v)=\sum_{j\in N^+_i} x_j$. We now calculate $\deg^+_{\overline{S}}(v)$; here, the constant $z_i$ comes into play. If $C_i\subseteq N^+_i$, $\deg^+_{\overline{S}\cap C_i}(v)=|C_i|-x_i-1=|C_i|-x_i-z_i$. Moreover, for $j\in N^+_i\setminus\{i\}$, $\deg^+_{\overline{S}\cap C_j}(v)=|C_j|-x_j$, so $\deg^+_{\overline{S}}(v)=\sum_{j\in N^+_i} (|C_j|-x_j)-z_i$. In the case of $i\notin N^+_i$, that is, $C_i$ is a negative clique or an independent set, then $\deg^+_{\overline{S}}(v)=\sum_{j\in N^+_i} (|C_j|-x_j)=\sum_{j\in N^+_i} (|C_j|-x_j)-z_i$.
\end{pf}
    
\end{toappendix}

For unsigned graphs, the distance to clique is a well-studied parameter. As we cannot expect to have cliques as a simple case for signed graphs, we rather propose the parameters \emph{distance to $k$-balanced complete graph} and \emph{distance to $k$-anti-balanced complete graph}. In both cases, technically speaking, we could consider the number~$k$ as a fixed constant or as another parameter. 

\begin{theorem}\label{thm:distance-parameters}
Let $\cal P$ be a problem on signed graphs that can be solved in \FPT-time when parameterized by neighborhood diversity. Then, $\cal P$ can also be solved  in \FPT-time when parameterized by any of the following parameters: \shortversion{1) vertex cover number of the underlying graph; 2) distance to $k$-balanced complete graph and $k$; 3) distance to $k$-anti-balanced complete graph and $k$.}\longversion{    \begin{enumerate}
        \item vertex cover number of the underlying graph;
        \item distance to $k$-balanced complete graph and $k$;
        \item distance to $k$-anti-balanced complete graph and $k$.
    \end{enumerate}}
\end{theorem}

\begin{pf} Let $G=(V,E^+,E^-)$ be a signed graph. 
For 1), we refer to \cite{ArrFFMQW2023}.
    2) Let $D\subseteq V$ be such that $|D|$ is the distance to $k$-balanced complete graph and $V-D= \bigcup_{i=1}^kV_i$, so that $(V_1,V_2,\dots,V_k)$ is the balanced partition of $G-D$. For each $v\in V_j$, $1\leq j\leq k$, its neighbors in $V-D$ are the same, and $v$ has at most $3^{|D|}$ different neighborhoods in~$D$, so that $\snd(G)\leq k\cdot 3^{|D|}+|D|$.
    3) Let $D'\subseteq V$ be such that $|D'|$ is the distance to $k$-anti-balanced complete graph. Analogously to 2), we have $\snd(G)\leq k\cdot3^{|D'|}+|D'|$.
\end{pf}

\shortversion{\noindent }Consequently, we have the \FPT results on a bunch of parameters\longversion{, which functionally bound on the signed neighborhood diversity}.
\begin{corollary}\label{cor:oa-nd}
    \textsc{Offensive Alliance} on signed graphs could be solved in \FPT-time when parameterized by any of the 
    parameters mentioned in \autoref{thm:distance-parameters}.
\end{corollary}

A natural research question is to find \FPT algorithms with better running times for these structural parameters. \longversion{Let $G=(V,E^+,E^-)$ be a signed graph. In this context, it is worth mentioning that, g}\shortversion{G}iven $G\shortversion{=(V,E^+,E^-)}$ and some\longversion{ integer} $\ell\geq 0$, it can be decided in \FPT-time if $\vc((V,E^+\cup E^-))\leq \ell$. As the distance to 1-balanced complete graph equals $\vc((V,\binom{V}{2}\setminus E^+))$ and the distance to 1-anti-balanced complete graph equals $\vc((V,\binom{V}{2}\setminus E^-))$, these parameters can be computed in \FPT-time, as well.
If $G$ is complete,\longversion{ i.e., $\binom{V}{2}=E^+\cup E^-$,} then (by~\cite{Har5354}) the distance to 2-balanced complete graph equals the odd cycle transversal number of $(V,E^-)$, which can be computed in \FPT-time by \cite{CygDLMNOPSW2016\longversion{,ReeSmiVet2004}}.\longversion{ A similar statement is true for 2-anti-balanced complete graphs.}
Further cases are still under investigation.
\longversion{For the case when the number~$k$ of clusters is not given, \cite[Theorem~2]{Dav67} gives a nice characterization: we need to delete at most~$\ell$ vertices from the given signed graph such that the resulting signed graph is complete and does not contain any cycle with exactly one negative edge.}

More generally speaking, very little research has been devoted to the parameterized complexity of problems on signed graphs. In particular, good structural parameters, genuinely designed for signed graphs, are lacking.

\longversion{\paragraph{Acknowledgements.} This work is supported by National Natural Science Foundation of China (CN) with No. 11971271; Natural Science Foundation of Shandong Province with No. ZR2019MA008. We are also grateful for the support of Zhidan Feng through China Scholarship Council.}

\bibliographystyle{plain}
\bibliography{ab,hen}

\end{document}